\documentclass[aps,prl,twocolumn,nofootinbib]{revtex4-1}
\usepackage{epsfig}
\usepackage[colorlinks,linkcolor=blue,anchorcolor=blue,citecolor=blue,urlcolor=blue,breaklinks=true]{hyperref}
\usepackage{graphics}
\usepackage{color}
\usepackage{graphicx}
\usepackage{amsmath}
\usepackage{amsfonts}
\usepackage{upgreek}
\usepackage{slashed}
\usepackage{color}

\begin{document}
\author{Wenkai Fan$^{1}$}\email[]{Email: 131242013@smail.nju.edu.cn}
\author{Xiaofeng Luo$^{2}$}\email[]{Email: xfluo@mail.ccnu.edu.cn}
\author{Hongshi Zong$^{3,4,5}$}\email[]{Email: zonghs@nju.edu.cn}
\affiliation{$^{1}$ Kuang Yaming Honors School, Nanjing University, Nanjing 210093, China}
\affiliation{$^{2}$ Key Laboratory of Quark \& Lepton Physics (MOE) and Institute of Particle Physics, Central China Normal University, Wuhan 430079, China}
\affiliation{$^{3}$ Department of Physics, Nanjing University, Nanjing 210093, China}
\affiliation{$^{4}$ Joint Center for Particle, Nuclear Physics and Cosmology, Nanjing 210093, China}
\affiliation{$^{5}$ State Key Laboratory of Theoretical Physics, Institute of Theoretical Physics, CAS, Beijing 100190, China}

\title{Identifying the presence of the critical end point in QCD phase diagram by higher order susceptibilities}
\begin{abstract}
 For the first time, we investigate susceptibilities of dense quark matter up to $8$th order using an effective model. Generally higher order susceptibilities will have more sign changes and larger magnitude, thus should give more information about the presence and location of the conjectured QCD critical end point (CEP). Two cases are studied, one with the CEP and one being crossover phase transition throughout the QCD phase diagram. It is found that a rapid crossover transition can also give similar sign pattern as the case with the CEP and yield large signal. We propose that using several kinds of ions for collision to search for a wider range on the QCD phase diagram may help.
\end{abstract}


\maketitle
Exploring the phase structure of strongly interacting nuclear matter is one of the main goals of heavy-ion collision experiments. Due to the asymptotic freedom of QCD, nuclear matter is expected to undergo a phase transition from a phase with hadrons as dominant degrees of freedom to a quark-gluon plasma (QGP) \cite{stephanov2005qcd}.
Lattice QCD calculations show that at small baryon chemical potential and high temperature, the transition is a smooth crossover \cite{fodor2002new}, whereas a first-order phase transition is expected at high baryon chemical region \cite{masayuki1989chiral,halasz1998phase,stephanov2006qcd,ejiri2008canonical}. The end point of this possible first-order phase boundary towards the crossover region is called the QCD critical end point (CEP). Due to the sign problem of lattice QCD, our current knowledge about this CEP are based on model calculations, but the CEP does now show up under all circumstances \cite{kitazawa2002chiral,mercado2011qcd,kohyama2015regularization}. So it is important to check if we can tell the existence of the CEP by experiment.

It has long been predicted that the fluctuations (susceptibilities) of baryon number and electric charge are sensitive to the phase transition. The experimental measurements of the fluctuations of conserved quantities have been performed in the beam energy scan (BES) program by the STAR and PHENIX experiments at the Relativistic Heavy-Ion Collider (RHIC). Interestingly, the STAR experiment observed a non-monotonic energy dependence of the fourth order ($\kappa\sigma^{2}$) net-proton fluctuations in the most central Au+Au collisions \cite{Luo201675,luo2017search}. Furthermore, this non-monotonic behavior cannot be described by various transport models \cite{xu2016cumulants,He2016296,Luo201675}. However, one important question remains. How can we tell that the non-monotonic behaving signal we found is caused by the CEP or just by a crossover transition? To investigate the contribution of the possible criticality physics to the conserved charges fluctuations, we adopt an effective quark model, the Nambu-Jona-Lasinio (NJL) model \cite{nambu1961dynamical,klevansky1992nambu}, to calculate the various fluctuations up to $8$th order in two cases, one with the CEP and one without. By comparing the difference between the two cases, one can see how critical behavior will influence the signal and whether current and forthcoming experiments should be able to tell the existence of the CEP.

Previous work has studied these quantities up to fourth order \cite{hatta2003universality,chen2015baryon,stephanov2009non,asakawa2009third,stephanov2011sign,chen2016robust,fan2016mapping}. We choose the NJL model as a representative since other effective models like the Polyakov-loop improved NJL model \cite{fu2010fluctuations,Cui:2013aba}, linear $\sigma$ model \cite{bowman2009critical}, the Polyakov-Quark-Meson (PQM) model \cite{friman2011fluctuations}, the Gross-Neveu (GN) model \cite{chen2015baryon} or Dyson-Schwinger equations \cite{qin2011phase,luecker2013critical,Zhao:2014oha,Cui:2015xta,Xu:2015vna,Cui:2016zqp} all share similar phase diagram or low-order susceptibilities with the NJL model and we are mainly interested in the qualitative behavior of the susceptibilities (the location of the CEP in the NJL model tends to higher $\mu_B$ and lower $T$ though).
The chemical potential of $u,d$ quarks are almost the same in experiments \cite{das2015chemical}, so we set them to be equal throughout the calculation. The chemical potential of the strange quark is smaller, but due to the large mass of $s$ quark, it does not vary the phase diagram much, thus having small influence on the susceptibilities \cite{fan2016mapping}. Throughout our calculation, we assume that the fire-ball is near thermal equilibrium at freeze-out, though critical slowing of dynamics would be important if the fire-ball passes the CEP \cite{berdnikov2000slowing,athanasiou2010using}. Additionally, changes in expansion dynamics and interactions that produce variations in particle spectra and acceptance independent of critical phenomena may blur the signal \cite{koch2010hadronic}.
\begin{figure*}[htbp]
  \centering
  \includegraphics[width=0.78\textwidth]{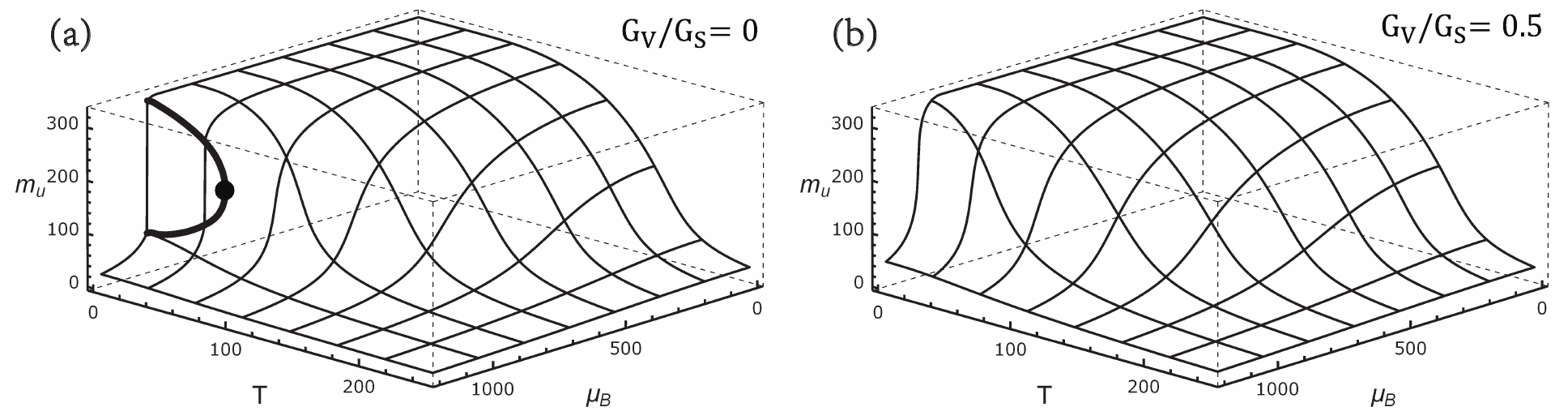}\\
  \caption{The phase diagram of the order parameter $m_u$ (the constituent up quark mass) in $MeV$. (a) $G_V=0$ case where a CEP is present. The thick line shows the location of the first-order phase transition (b) $G_V=0.5G_S$ case where there is no critical behavior but crossover transition}\label{fig:quark_mass}
\end{figure*}

The lagrangian density we adopt is the 3-flavor NJL model with scalar and vector interactions, along with the t'Hooft interaction which breaks the $U(1)_A$ symmetry:
\begin{equation}\label{equ:original_lagrangian}
\begin{aligned}
\mathcal{L}=&\overline{\psi}(i\slashed \partial-m)\psi+G_S[(\overline{\psi}\lambda_i\psi)^2+(\overline{\psi}i \gamma_5 \lambda_i\psi)^2]\\
&-G_V[(\overline{\psi}\gamma_{\mu} \lambda_i\psi)^2+(\overline{\psi}\gamma_{\mu} \gamma_5 \lambda_i\psi)^2]\\
&-K({det[\overline{\psi}(1+\gamma_5)\psi]+det[\overline{\psi}(1-\gamma_5)\psi]})
\end{aligned}
\end{equation}
The model's parameters are taken from Ref.~\cite{hatsuda1987effects}. The bare quark masses are $m_{u0}=m_{d0}=5MeV,m_{s0}=136MeV$. The 3-momentum cutoff $\Lambda=631MeV$, $G_S=1.83/\Lambda^2$ and $K=9.29/\Lambda^5$. After mean-field approximation, the following equations hold:
\begin{equation}\label{equ:0th order}
\left\{
\begin{aligned}
&m_i=m_{i0}-4G_S \langle\overline{q}_i q_i\rangle+2K \langle\overline{q}_m q_m\rangle \langle\overline{q}_n q_n\rangle (i\neq m \neq n)\\
&\mu_i=\mu_{i0}-4 G_V\langle q_i^{\dag}q_i\rangle\\
\end{aligned}
\right.
\end{equation}
where $\langle\Theta\rangle=\frac{Tr (\Theta e^{-\beta(\mathcal{H}-\mu_i \mathcal{N}_i)})}{Tr( e^{-\beta(\mathcal{H}-\mu_i\mathcal{N}_i)})}$ being the grand canonical ensemble average, and $i=u,d,s$.

\begin{figure*}[htbp]
  \centering
  \includegraphics[width=0.84\textwidth]{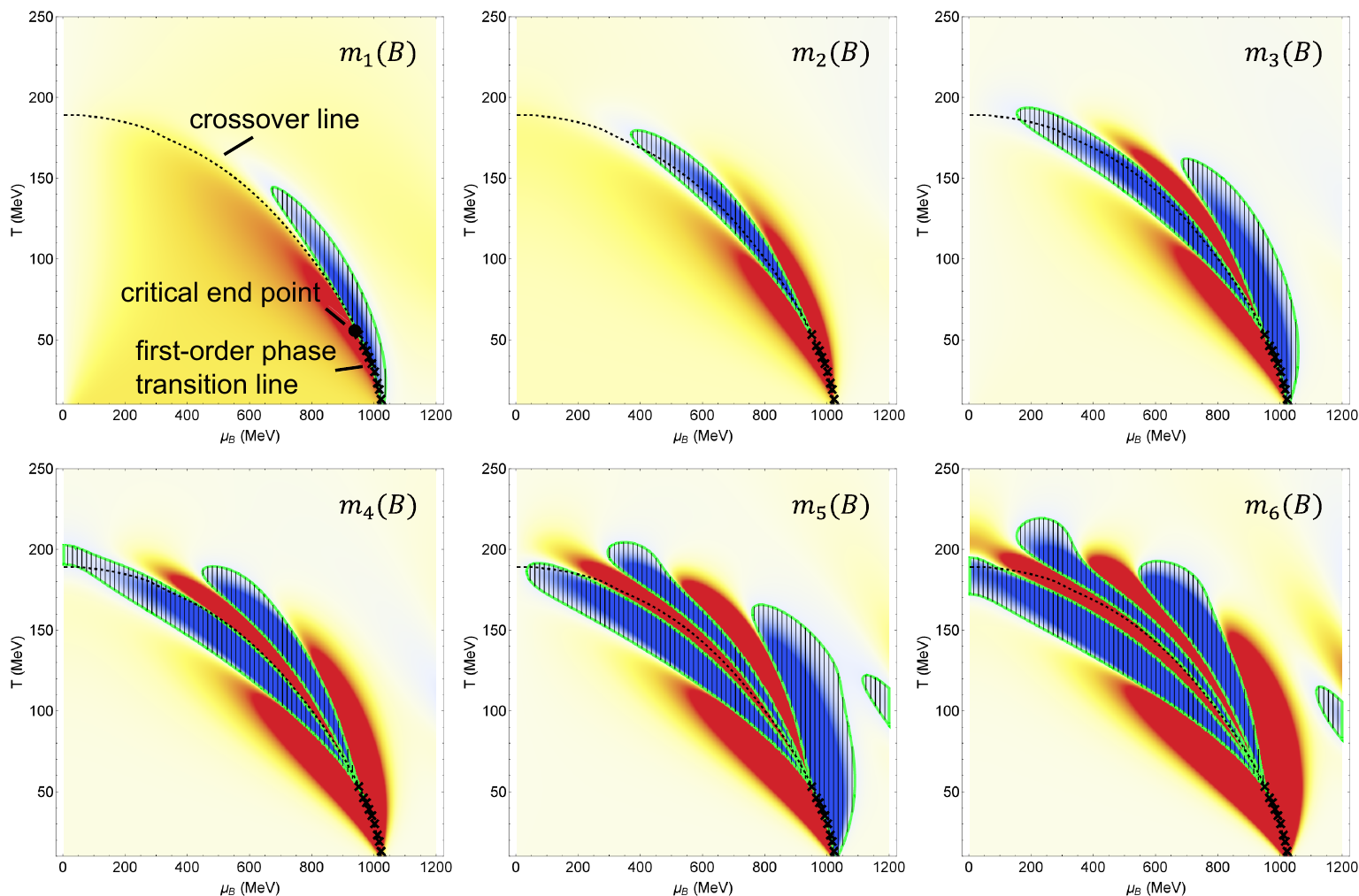}\\
  \caption{Sign of $m_n$ of baryon number for the $G_V=0$ case. Red region represents positive value while blue zone represents negative value. The dashed line is the crossover line while the crosses represents the first--order phase transition curve. The negative region are also enclosed by solid green line and filled with stripes for illustration purpose.}\label{fig:m2d_B_Gv=0}
\end{figure*}

\begin{figure*}[htbp]
  \centering
  \includegraphics[width=0.84\textwidth]{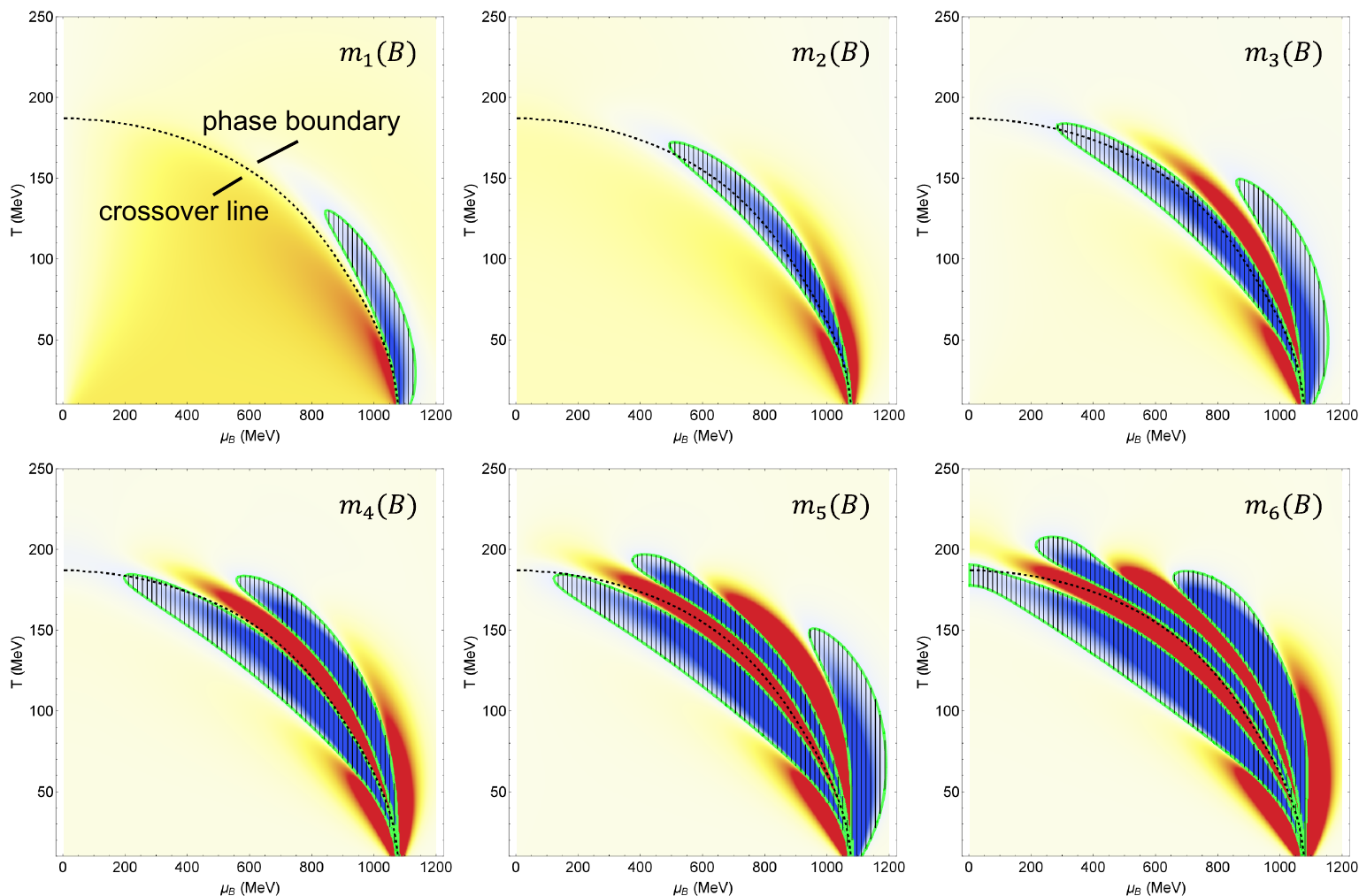}\\
  \caption{Sign of $m_n$ of baryon number for the $G_V=0.5G_S$ case. Red region represents positive value while blue zone represents negative value. The dashed line is the crossover line while the crosses represents the first--order phase transition curve. The negative region are also enclosed by solid green line and filled with stripes for illustration purpose.}\label{fig:m2d_B_Gv=0.5}
\end{figure*}

The various susceptibilities are defined as:
\begin{equation}
\begin{aligned}
&\frac{\partial\langle q_i^{\dag}q_i\rangle}{\partial \mu_j}=\chi_{ij},\frac{\partial^2\langle q_i^{\dag}q_i\rangle}{\partial \mu_j\partial \mu_k}=\chi_{ijk},\frac{\partial^3\langle q_i^{\dag}q_i\rangle}{\partial \mu_j\partial \mu_k\partial \mu_p}=\chi_{ijkp}
\end{aligned}
\end{equation}
Furthermore, we change the base from $\{u,d,s\}$ at quark level to the conserved charges $\{B,Q,S\}$ by using:
\begin{equation}\label{equ:basechange}
\left\{
\begin{aligned}
&\mu_u=\frac{1}{3}(\mu_B+2\mu_Q)\\
&\mu_d=\frac{1}{3}(\mu_B-\mu_Q)\\
&\mu_s=\frac{1}{3}(\mu_B-\mu_Q-3\mu_S)
\end{aligned}
\right.
\end{equation}

We consider two cases: one with $G_V=0$ which has a CEP, another one with $G_V=0.5G_S$ (given by renormalization-group analysis \cite{evans1999effective,schafer1999high}) which is crossover transition throughout the phase diagram \cite{kitazawa2002chiral}. As can be seen from Fig.~\ref{fig:quark_mass}, the two cases have very similar behavior at low $\mu_B$.

In order to relate our calculation with experiments and other model calculation, we consider the following ratios defined as:
\begin{equation}\label{}
  m_n(X)=\frac{T^n\chi_{X}^{(n+2)}}{\chi_{X}^{(2)}}, n=1,2,3...
\end{equation}
where $x=B, Q, S$. These ratios are then independent of the volume of the system.  The signs of these ratios of baryon number are shown in Fig.~\ref{fig:m2d_B_Gv=0} and ~\ref{fig:m2d_B_Gv=0.5}. Red regions are of positive value, and blue regions are of negative value. The yellow regions represent values very close to $0$.

Within the phase boundary, there is really not much difference of the signs of the signals (the moments). The negative regions in Fig.~\ref{fig:m2d_B_Gv=0} (a) and (b) are the same as predicted in Refs.~\cite{asakawa2009third} and \cite{stephanov2011sign} by means of universal analysis. However, similar sign patterns appear also in the case with no CEP (see Fig.~\ref{fig:m2d_B_Gv=0.5}). If we measure points away from the phase boundary, we may not be able to tell whether the CEP is present by only analyzing the sign of the signals. The most significant difference between the two cases lies at the low $T$ (large $\mu_B$) part of the phase boundary.
Across the first-order phase transition line (crosses in Fig.~\ref{fig:m2d_B_Gv=0}), the signal changes sign for $0$ (even order) or $1$ (odd order) time. Within the $G_V=0.5G_S$ case, however, the signal changes sign more and more times across the crossover line as the order becomes higher. If we are able to measure enough points across this part of the phase boundary, we may be able to tell whether the phase transition is first order or crossover. This is a very important conclusion of our analysis.

\begin{figure*}[htbp]
  \centering
  \includegraphics[width=1\textwidth]{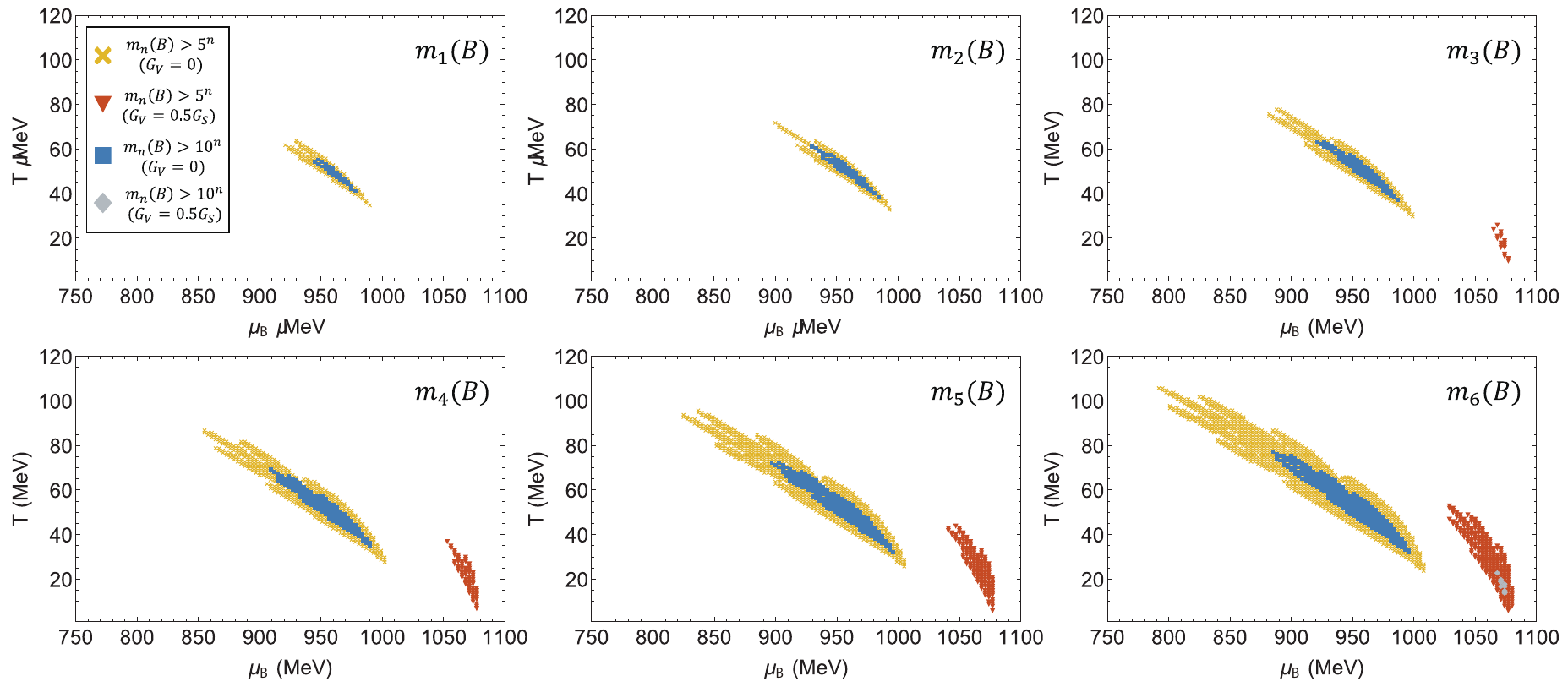}\\
  \caption{Region where the magnitude of $m_n(B)>5^n$ (yellow crosses and red triangles) or $10^n$ (blue squares and gray diamonds) for the two cases. The left regions belong to the $G_V=0$ case and the right regions belong to $G_V=0.5G_S$ case.}\label{fig:region_compare}
\end{figure*}

Next, we want to study the magnitude of the signals on the phase diagram. In Fig.~\ref{fig:region_compare}, we plot the region where the magnitude of $m_n(B)$ being greater than $5^n$ or $10^n$ for both the two cases. The case with the CEP has a large area where large signals are expected while the case with no CEP has only a small area. Besides, due to criticality, the magnitude of signals of the case $G_V=0$ can be even larger if we become close enough to the CEP (in Fig.~\ref{fig:region_compare}, the region where $m_n(B)>10^n$ is still sizable), where the magnitude of signals of the case $G_V=0.5G_S$ is limited. By only analyzing the magnitude of the signal in experiments, we can not yet tell whether the CEP is present. But if the magnitude of the signals are very large, the chances are that there is a CEP or the crossover phase transition is very rapid.

In this letter, we mainly discussed about three aspects by considering two cases with and without the critical end point. First of all, higher order fluctuations will carry more information about the phase transition. The flip of sign will indicate the location of the phase transition and the magnitude of higher-order signals are generally larger. It should be very meaningful to measure higher-order fluctuations. Secondly, we find that the case with no critical behavior can give similar sign pattern in the vicinity of the phase transition. The two cases considered in this letter have similar behavior at low chemical potential and are not in conflict with lattice simulation \cite{allton2005thermodynamics}, so they are both possible candidate for the real QCD phase diagram. Also, their sign pattern and magnitude both agree with current experiment data qualitatively \cite{Luo201675} (the comparison between the case with CEP and experiment data is done in Ref.~\cite{fan2016mapping}, and the case with no CEP does not differ much from the case with the CEP at this order). By only analyzing the sign of current and future experiment signal, we may not be able to tell whether the CEP is present. We need to move across the phase boundary (the possible first-order transition line) for more information. Thirdly, the magnitude of the fluctuations of the two cases differ a lot. The case with the CEP will give very large signals if we can come close enough to the CEP. However, a rapid crossover transition can also yield sizable signals. Further investigation is still needed. It should be meaningful to use different kinds of ions (different freeze-out curves) for collision in order to search for a wide region on the phase diagram \cite{gazdzicki2015search,vovchenko2016hadron}. If there is a CEP and all the data points we measured are inside the phase boundary, we should expect an rapid increase of the magnitude of the signals as we approach the phase boundary. If we pass through the CEP, we can locate the CEP within the nearest two freeze-out lines. In this way, we can reduce the uncertainty of the existence and location of the CEP.

\acknowledgments
The work is supported in part by the National Natural Science Foundation of China (under Grants No. 11475085, No. 11535005, No.11690030, No. 11575069, and No. 11221504), and the MoST of China 973-Project No. 2015CB856901.

%


\bibliographystyle{apsrev4-1}
\bibliography{sus}

\end{document}